\documentclass[twocolumn,aps,floatfix,superscriptaddress]{revtex4}
\pdfoutput=1 
\usepackage{amsmath,amssymb,eucal,graphicx,float,epstopdf}
\begin{document}
\title{Monotonicity in the averaging process}

\author{E.~Ben-Naim}
\affiliation{Theoretical Division and Center for Nonlinear Studies,
Los Alamos National Laboratory, Los Alamos, New Mexico 87545, USA}
\author{P.~L.~Krapivsky}
\affiliation{Department of Physics, Boston University, Boston, Massachusetts 02215, USA}
\affiliation{Santa Fe Institute, 1399 Hyde Park Road, Santa Fe, NM 87501, USA}
\affiliation{Skolkovo Institute of Science and Technology, 143026 Moscow, Russia}
\begin{abstract}
We investigate an averaging process that describes how interacting
agents approach consensus through binary interactions.  In each
elementary step, two agents are selected at random and they reach
compromise by adopting their opinion average.  We show that the
fraction of agents with a monotonically decreasing opinion decays as
$e^{-\alpha t}$, and that the exponent
$\alpha=\tfrac{1}{2}-\tfrac{1+\ln \ln 2}{4\ln 2}$ is selected as the
extremum from a continuous spectrum of possible values.  The opinion
distribution of monotonic agents is asymmetric, and it becomes
self-similar at large times.  Furthermore, the tails of the opinion
distribution are algebraic, and they are characterized by two distinct
and nontrivial exponents.  We also explore statistical properties of
agents with an opinion strictly above average.
\end{abstract}

\maketitle

\section{Introduction}

The averaging process models a system of agents who reach agreement
via compromise.  The system consists of $N$ agents.  In each step, two
agents, say $i$ and $j$, are chosen at random and their opinions $v_i$
and $v_j$ are replaced by the average
\begin{equation}
\label{averaging}
(v_i,v_j) \to \left(\frac{v_i+v_j}{2},\frac{v_i+v_j}{2}\right)\,.
\end{equation}
As this elementary step is repeated, the system moves closer and
closer toward consensus where all agents have the same
opinion.

This averaging process has been studied by statistical physicists
\cite{CFL}, applied probabilists \cite{AL}, computer scientists
\cite{OT}, and social scientists \cite{HK} with applications ranging
from opinion dynamics \cite{flpr} to communication algorithms for
computer and sensor networks \cite{sdj,pbb} to linguistics
\cite{dcll}. For the pure averaging process \eqref{averaging}, the
system approaches perfect consensus.  However, when interactions are
restricted to agents with sufficiently close opinions, the system
bifurcates into groups \cite{HK,WDAN,BKR,JL,OH,BS}, with all agents
within the same group sharing the same opinion.  Another
generalization of the averaging process \eqref{averaging} involves
partial averaging where the opinion difference is reduced by a fixed
multiplicative factor in each interaction.  Partial averaging is
equivalent to inelastic collisions, and it has been used to model
freely evolving and driven inelastic gases
\cite{BK00,BMP02,BK04,KB02,EB02,ADL,PDSR,GS,NG,book} that satisfy the
Maxwell model rules \cite{M67}.

In this study, we analyze the pure averaging process
\eqref{averaging}. The system approaches perfect consensus and the
difference between the typical opinion and the consensus opinion
decays exponentially with time \cite{BK00}.  While the typical opinion
follows a unidirectional path en route to consensus, an individual
opinion may increase or decrease due to fluctuations. We focus on
agents with a monotonically decreasing opinion (similar behavior is
exhibited by agents with a monotonically increasing opinion).

We find that the fraction $M(t)$ of {\em monotonic} agents decays
exponentially with time,
\begin{equation}
\label{Mt}
M(t)\simeq A\,e^{-\alpha t}\,,
\end{equation}
in the long-time limit (see Fig.~\ref{fig-Mt}).  Our main result is
that the exponent $\alpha$ is nontrivial
\begin{equation}
\label{alpha}
\alpha=\frac{1}{2} - \frac{1+\ln \ln 2}{4\ln 2}=0.271517\ldots\,.
\end{equation}
The opinion distribution of monotonic agents becomes self-similar at
sufficiently large times. Further, this distribution has two algebraic
tails that are characterized by two different exponents.  These
features of the opinion distribution enable us to determine the
exponent $\alpha$ which is selected as the extremal value from a
spectrum of possible values.

The rest of this paper is organized as follows. We begin with a brief
overview of the averaging process in Sec.~\ref{sec:averaging}. In
Sec.~\ref{sec:monotonic}, we consider monotonic agents with an opinion
that only decreases with time. We obtain the exponent \eqref{alpha},
and also study the opinion distribution of monotonic agents.  In
Sec.~\ref{sec:positive}, we investigate the behavior of agents with
strictly positive opinion.  In Sec.~\ref{sec:summary}, we summarize
our findings and discuss possible avenues for future work. The
generalization of the pure averaging process \eqref{averaging} to
partial averaging is outlined in Appendix~\ref{ap:partial}, and
corrections to the leading asymptotic behavior \eqref{Mt} are
discussed in appendix~\ref{ap:corr}.

\section{The averaging Process}
\label{sec:averaging}

In the averaging process, there are $N$ interacting agents. In each
averaging event, two agents are selected at random, and their opinions
evolve according to \eqref{averaging}. This pairwise interaction is
repeated indefinitely, and time is augmented by $2/N$ after each
interaction, so that agents experience one averaging event per unit
time.  The distribution $F(v,t)$ of agents with opinion $v$ at time
$t$ satisfies the {\it nonlinear} rate equation \cite{BK00}
\begin{equation}
\label{Fvt-eq}
\frac{\partial F(v,t)}{\partial t} = 
2\int_{-\infty}^\infty du\,F(u,t)F(2v-u,t)-F(v,t)\,,
\end{equation}
in the limit $N\to \infty$. The convolution term reflects the binary
nature of \eqref{averaging}, and the linear loss rate reflects that
agents participate in one interaction per unit time. The averaging
process \eqref{averaging} conserves the number of particles and the
total opinion, and consequently, the rate equation \eqref{Fvt-eq}
conserves the two lowest moments of the opinion distribution: the
normalized distribution, $\int dv\, F(v,t)=1$, and the average
opinion, \hbox{$\langle v(t)\rangle=\int dv\,v\, F(v,t)={\rm const}$}.

The averaging process \eqref{averaging} is invariant under translation
\hbox{$v\to v+{\rm const.}$}, as well as dilation, \hbox{$v\to {\rm
    const.}\times v$}. Hence, without loss of generality, we may
consider initial distributions with zero average, $\langle
v(0)\rangle=0$, and unit variance, $\langle v^2(0)\rangle=1$
\cite{finite}.  Furthermore, we restrict our attention to symmetric
distributions as Eq.~\eqref{Fvt-eq} implies $F(v,t)=F(-v,t)$ when
$F(v,0)=F(-v,0)$.

The system approaches consensus, \hbox{$F(v,t)\to \delta(v)$}, as all
agents acquire the average initial opinion in the long-time limit. The
second moment \hbox{$\langle v^2(t)\rangle=\int_{-\infty}^\infty
  dv\,v^2 F(v,t)$} quantifies the distance between the typical opinion
and the consensus opinion.  This quantity decays exponentially with
time \cite{BK00}
\begin{equation}
\label{M2}
\langle v^2(t)\rangle = e^{-t/2}\,,
\end{equation}
as follows from Eq.~\eqref{Fvt-eq}.  Moreover, the distribution
$F(v,t)$ becomes self-similar in the long-time limit, and the second
moment \eqref{M2} sets the scale for the typical opinion. In
particular, the distribution $F(v,t)$ adheres to the scaling form
\cite{BMP02}
\begin{equation}
\label{Fvt-scaling}
F(v,t) = e^{t/4}\mathcal{F}(V), \quad {\rm with} \quad
\mathcal{F}(V) = \frac{2}{\pi}\,\frac{1}{(1+V^2)^2}\,,
\end{equation}
and the scaling variable $V=v\,e^{t/4}$. This scaling behavior holds
in the limits \hbox{$t\to\infty$} and \hbox{$v\to 0$}. Also, the
scaling function is normalized $\int_{-\infty}^{\infty}
dV\mathcal{F}(V)=1 $.

\section{Monotonicity}
\label{sec:monotonic}

The ultimate opinion of every agent vanishes, $v_i\to 0$ as $t\to
\infty$. Further, according to Eqs.~\eqref{M2}--\eqref{Fvt-scaling},
the typical magnitude of the opinion $|v|$ decreases monotonically,
$|v|\sim e^{-t/4}$. In this study, we focus on agents with a
monotonically decreasing opinion.  At time $t$, we refer to agents
with opinion that satisfies the inequality $v_i(t_1)\geq v_i(t_2)$ for
all $t_1\leq t_2\leq t$ as {\it monotonic} agents.  In the context of
opinion dynamics, monotonic agents change their opinion only in one
direction, say toward the left of the political spectrum
only. According to \eqref{averaging}, monotonic agents interact only
with agents who have a smaller opinion.  We stress that monotonic
agents have an opinion that strictly decreases with time, but the sign
of the opinion is not constrained. In particular, monotonic agents may
start with a positive opinion and end up with a negative one.

\begin{figure}[t]
\begin{center}
\includegraphics[width=0.45\textwidth]{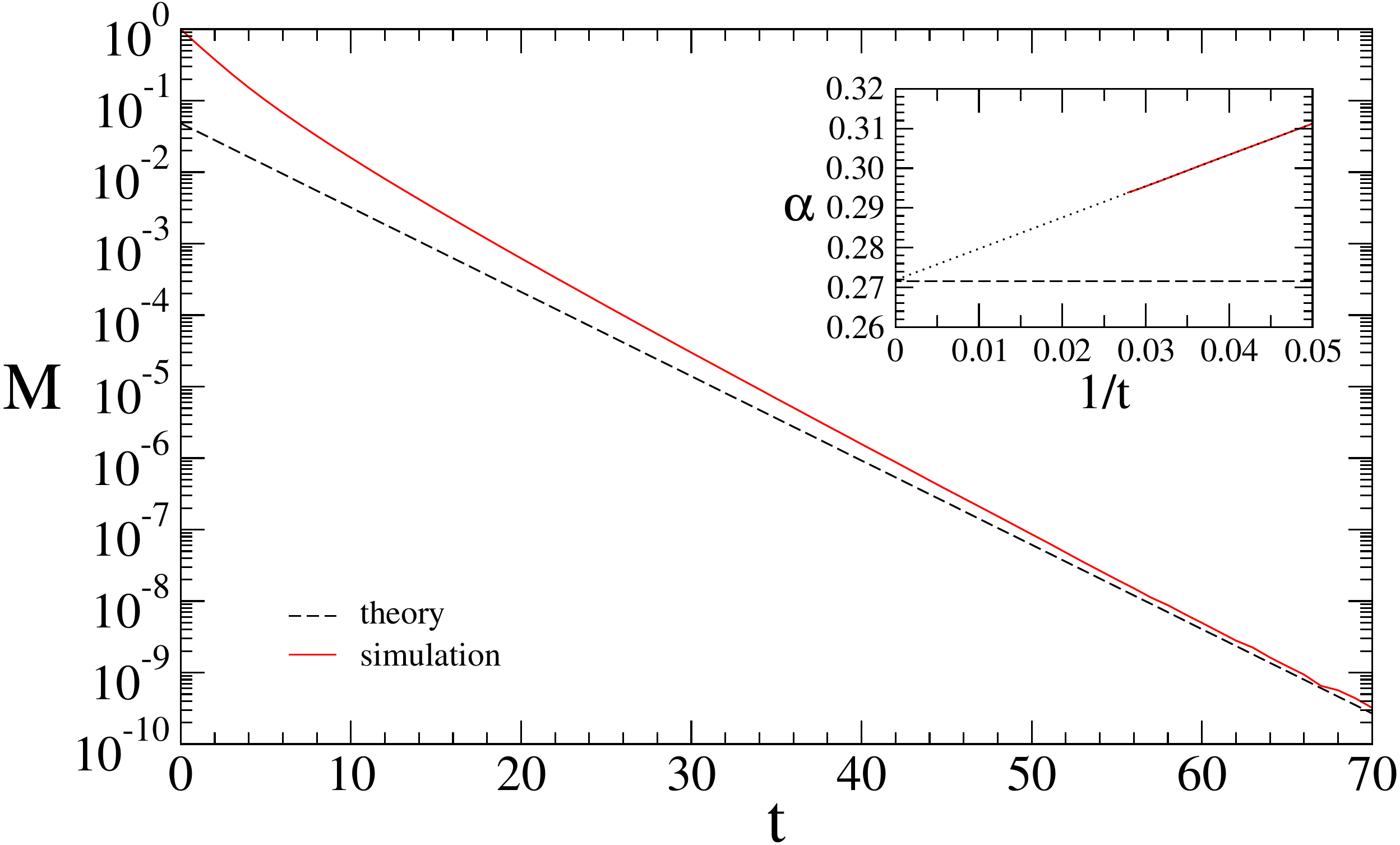}
\caption{The fraction of monotonic agents $M(t)$ versus time $t$. 
  A dashed line with slope
  given by Eq.~\eqref{Mt} is displayed for reference.  The inset shows
  that the local slope \hbox{$\alpha(t) \equiv -d\ln M /\ln t$} varies
  linearly with $1/t$. The dotted line shows a linear fit that
  yields the estimate $\alpha=0.272\pm 0.003$.}
\label{fig-Mt}
\end{center}
\end{figure}

We denote by $M(t)$ the fraction of monotonic agents at time $t$, and
by $M(v,t)$ the density of such agents with opinion $v$. Of course,
$M(t)=\int dv M(v,t)$.  By symmetry, the dual fraction of agents with
monotonically increasing opinion equals $M(t)$, and their density is
given by $M(-v,t)$.  The density $M(v,t)$ is coupled to the total
density $F(v,t)$, and it satisfies the {\em linear} rate equation
\begin{equation}
\label{Mvt-eq}
\frac{\partial M(v,t)}{\partial t} = 
2\int_v^\infty du\,M(u,t)F(2v-u,t)-M(v,t)\,.
\end{equation}
The loss term reflects that on average, each agent experiences one
interaction per unit time. The gain term in Eq.~\eqref{Mvt-eq}
resembles the gain term in Eq.~\eqref{Fvt-eq}, but the lower limit of
integration ensures that the opinion of a monotonic agent may only
decrease.

By integrating the master equation \eqref{Mvt-eq} over all opinions,
we find that the fraction $M(t)$ decreases with time according to the
rate equation
\begin{equation}
\label{Mt-eq}
\frac{d M(t)}{dt} = -\int_{-\infty}^\infty du\,M(u,t)\int_u^{\infty} dv\, F(v,t)\,.
\end{equation}
This evolution equation reflects that monotonic agents may only
interact with agents having smaller opinions. The fraction $M(t)$ has
two bounds, \hbox{$e^{-t}\leq M(t)\leq 1$}. The upper bound follows
from the inequality \hbox{$M(v,t)\leq F(v,t)$}. The lower bound
reflects that agents experiencing zero interactions are necessarily
monotonic. Since agents interact once per unit time, the overall
density $N(t)$ of noninteracting agents decays exponentially with
time, $N(t)=e^{-t}$.  The bounds $\alpha=0$ and $\alpha=1$ are
realized in limiting cases of the partial averaging process, as
discussed in Appendix~\ref{ap:partial}.

When the opinion $v$ is sufficiently large, the dominant contribution
to the integral in \eqref{Mvt-eq} comes from the vicinity of
$u=2v$. Using the normalization $\int_{-\infty}^\infty dv\, F(v,t)=1$
we arrive at the linear equation
\begin{equation}
\label{cascade-eq}
\frac{\partial M(v,t)}{\partial t}=2M(2v,t)-M(v,t)\,,
\end{equation}
that holds for sufficiently large $v$.  The derivation of
Eq.~\eqref{cascade-eq} relies on the fact that the distribution
$F(v,t)$ is normalized, but remarkably, the precise form of that
distribution is not utilized.  Indeed, as follows directly from
\eqref{averaging}, the outcome of interactions involving an agent with
a very large opinion $v$ is not affected by the opinion of the
interaction counterpart.  Consequently, large opinions are reduced by
a factor $2$ with each interaction, thereby leading to a simple
multiplicative process \cite{bm}
\begin{equation}
\label{cascade}
v\to v/2 \to v/4 \cdots.
\end{equation}
Equation \eqref{cascade-eq} merely reflects this multiplicative
process.

\begin{figure}[t]
\begin{center}
\includegraphics[width=0.45\textwidth]{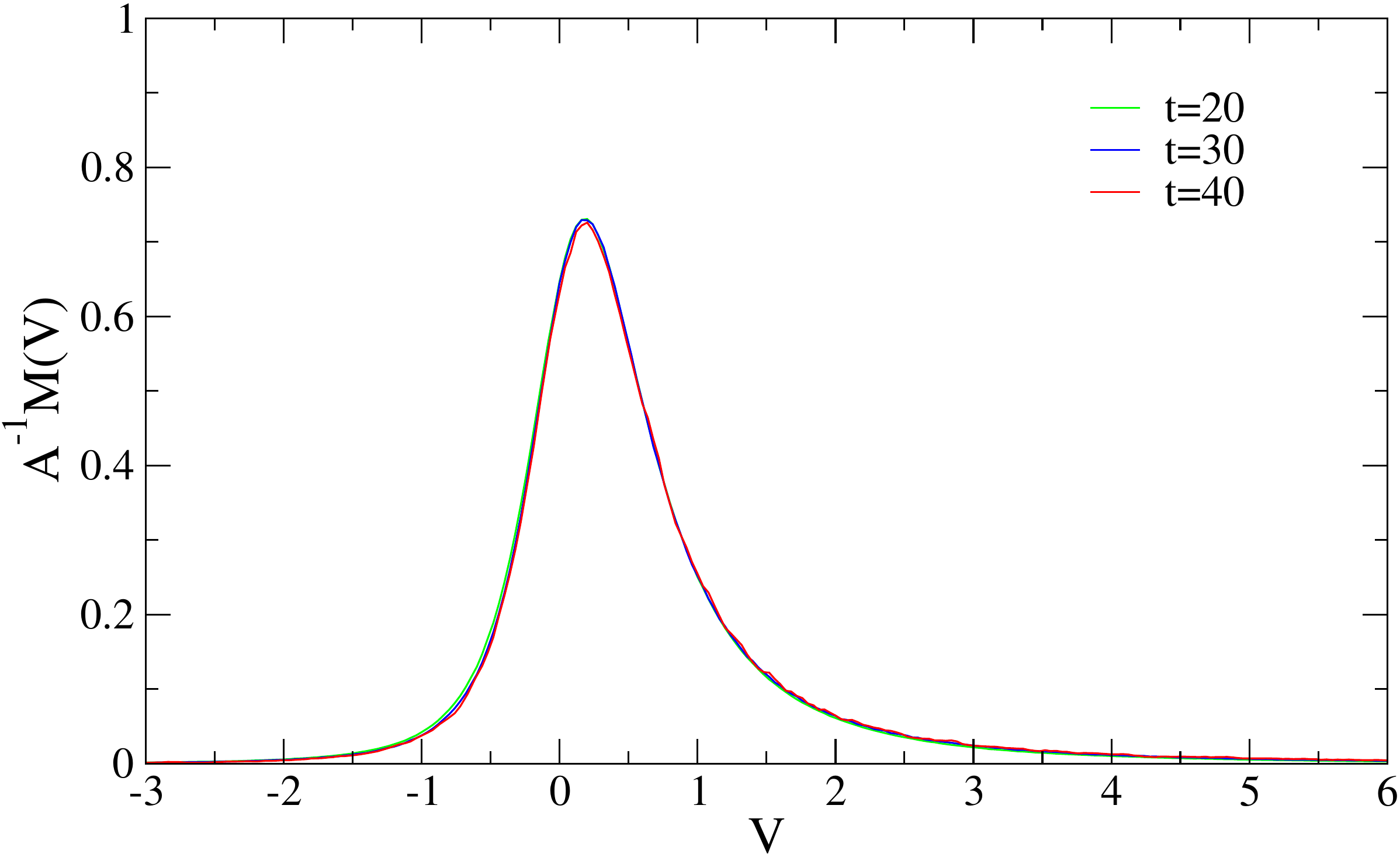}
\caption{The normalized scaling function $A^{-1}\mathcal{M}(V)$ versus
  the scaling variable $V$. This quantity is the probability
  distribution function of the scaled opinion $V$, as follows from
  Eqs.~\eqref{Mt} and \eqref{Mvt-scaling}.}
\label{fig-MV}
  \end{center}
\end{figure}

In the long-time limit, the density $M(v,t)$ of monotonic agents
adheres to the scaling form
\begin{equation}
\label{Mvt-scaling}
M(v,t) \simeq e^{-\alpha t}e^{t/4}\,\mathcal{M}(V)
\end{equation}
with the same scaling variable $V=ve^{t/4}$. By integrating
\eqref{Mvt-scaling}, we obtain the exponential decay \eqref{Mt} with
\hbox{$A= \int_{-\infty}^\infty dV\,\mathcal{M}(V)$}. Figure
\ref{fig-MV} convincingly shows that the scale $v\sim t^{-1/4}$ also
characterizes the opinion distribution of monotonic agents.

\begin{figure}[t]
\begin{center}
\includegraphics[width=0.45\textwidth]{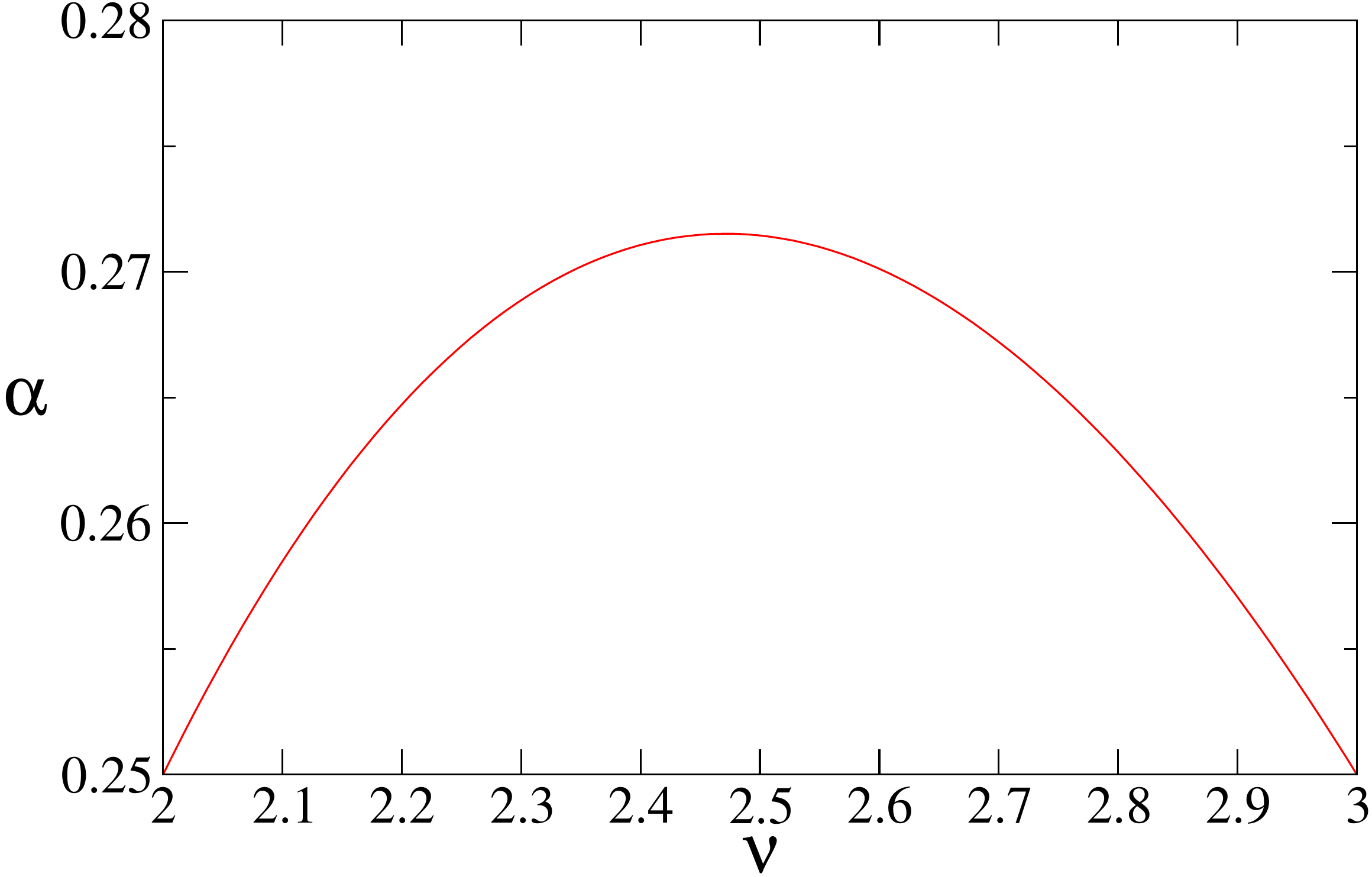}
\caption{The exponent $\alpha$ in \eqref{alpha-nu} versus $\nu$. The
  maximal value $\alpha=0.271517$ occurs at $\nu= 2.47123$. }
\label{fig:alpha-nu}
\end{center}
\end{figure}

By substituting the scaling form \eqref{Mvt-scaling} into
\eqref{cascade-eq}, we find that the scaling function $\mathcal{M}(V)$
satisfies 
\begin{equation}
\label{MV-tail-eq}
\frac{V}{4}\,\frac{d\mathcal{M}(V)}{dV} 
+ \left(\frac{5}{4}-\alpha\right) \mathcal{M}(V)
= 2\mathcal{M}(2V)
\end{equation}
when $V\gg 1$.  This difference-differential equation admits an
algebraic solution, $\mathcal{M}(V) \sim V^{-\nu}$, with
\begin{equation}
\label{alpha-nu}
\alpha=\frac{5-\nu}{4}-2^{1-\nu}\,.
\end{equation}
The right-hand side of \eqref{alpha-nu} is bounded from above, see
Fig.~\ref{fig:alpha-nu}.  The maximal value quoted in \eqref{alpha}
occurs at 
\begin{equation}
\label{nu-star}
\nu = 3+\frac{\ln \ln 2}{\ln 2}=2.471233\ldots\,.
\end{equation}
Thus, the algebraic tail for $V\gg 1$ yields the upper bound
$\alpha\leq \alpha_*$. We postulate that this extremal value is
realized, $\alpha=\alpha_*$, for the averaging process.  Our numerical
simulations give the estimate $\alpha=0.272\pm 0.003$ and hence
support this theoretical prediction (see the inset to
Fig.~\ref{fig-Mt} and also, Appendix B).  We note that our assumption
that the behavior in the neighborhood of $u=2v$ dominates the integral
in Eq.~\eqref{Mvt-eq} is consistent with the fact that the tail
\hbox{$M(v) \sim v^{-\nu_*}$} is shallower than the tail $F(v)\sim
v^{-4}$.  The selection of the extremal value emerging from the
dispersion-like relation \eqref{alpha-nu} very much resembles the
selection of the extremal propagation velocity in traveling waves that
are governed by partial differential equations in deterministic
\cite{KPP,Fisher,Bramson,Saarloos,Saarloos-Rev} and 
stochastic systems \cite{BD97,KM00,KM02,BK03,BD06}.

Next, we substitute the scaling form \eqref{Mvt-scaling} into the
master equation \eqref{Mvt-eq} and arrive at the linear
integro-differential equation 
\begin{equation}
\label{MV-eq}
\frac{V}{4}\,\frac{d\mathcal{M}}{dV} + \left(\frac{5}{4}-\alpha\right) \mathcal{M}
= 2\int_V^\infty dU\,\mathcal{M}(U)\,\mathcal{F}(2V-U)
\end{equation}
with $\mathcal{F}(V)$ given in Eq.~\eqref{Fvt-scaling}. We stress that
this equation governs the scaling equation $\mathcal{M}(V)$ for all
values of $V$, in contrast with Eq.~\eqref{MV-tail-eq} that applies
only when \hbox{$V\gg 1$}.  The parameter $\alpha$ is an eigenvalue of this
equation, and in principle, a solution for the eigenvalue $\alpha$
requires a solution for the entire eigenfunction $\mathcal{M}(V)$.
However, in our particular problem, extreme-value analysis
suffices.

To understand the behavior when $V\ll -1$, we introduce the change of
variables $U=2V-W$, and thereby recast the right-hand side of
Eq.~\eqref{MV-eq} into \hbox{$2\int_{-\infty}^V\!
  dW\,\mathcal{M}(2V-W)\,\mathcal{F}(W)$}\,.  In the limit \hbox{$V\to
-\infty$}, this integral is negligible and Eq.~\eqref{MV-eq} simplifies
to $V\mathcal{M}' + (5-4\alpha) \mathcal{M}=0$. Therefore, there is a
second algebraic tail, $M(V)\sim (-V)^{-(5-4\alpha)}$ when \hbox{$V\ll
  -1$}.

Thus, the scaling function $\mathcal{M}(V)$ has two distinct algebraic
tails
\begin{equation}
\label{MV-tails}
\mathcal{M}(V)\sim
\begin{cases}
(-V)^{-(5-4\alpha)} & V\ll -1\,, \\
V^{-\nu}             & V\gg 1\,.
\end{cases}
\end{equation}
The asymmetry of $\mathcal{M}(V)$ is reflected by the inequality
$\nu<5-4\alpha$ as $5-4\alpha=3.913932$.  Interestingly, both
tails of the scaling function $\mathcal{M}(V)$ are shallower than the
tails of the scaling function $\mathcal{F}(V)$ as both $\nu<4$ and
\hbox{$5-4\alpha<4$}.  Of course, the inequality \hbox{$M(v,t)\leq F(v,t)$}
holds for all $v$.  By comparing the two densities $e^{-\alpha t}
\mathcal{M}(V_\pm)\sim \mathcal{F}(V_\pm)$, we find that
\eqref{Fvt-scaling} and \eqref{Mvt-scaling} hold simultaneously in the
scaling region $V_{-} \ll V \ll V_{+}$.  Both scales $V_+$ and $V_-$ grow
exponentially with time, and hence, the two scaling forms
\eqref{Fvt-scaling} and \eqref{Mvt-scaling} hold over an exponentially
growing range of scaled opinions.

Another consequence of the asymmetry of
$\mathcal{M}(V)$ is that the majority of monotonic agents have a
positive opinion. This fraction saturates at a finite value
\begin{equation}
\label{M+}
M_+=\int_0^\infty dV\, \mathcal{M}(V)\,.
\end{equation}
Using numerical simulations we find $M_+=0.74\pm 0.01$.  The
complementary fraction $M_-$ of monotonic agents with negative
opinion, $M_-=1-M_+$, is roughly three times smaller than the fraction
of monotonic agents with a positive opinion.

Our Monte Carlo simulations utilize a straightforward implementation
of the averaging process. Initially, there are $N$ agents whose
opinions are drawn from a uniform distribution: $F(v,0)=1$ for
$|v|<1/2$ and $F(v,0)=0$ otherwise.  We stress that our main results
are independent of the shape of the initial distribution: the same
scaling functions $\mathcal{F}(V)$ and $\mathcal{M}(V)$ are realized
for compact distributions as well as non-compact distributions
\cite{finite}. In each simulation step, two agents are selected at
random, and their opinions are updated according to the averaging rule
\eqref{averaging}. Time is augmented by $2/N$ subsequently.  We
maintain a counter for the number of monotonic agents, and whenever an
agent interacts for the first time with an agent having a larger
opinion, the counter decreases by one. The simulation results shown
throughout this paper represent an average over $10^3$ independent
Monte Carlo runs in a system of size $N=10^8$.

We made the following choices to optimize the simulations: (i) we
adjust the average initial velocity to zero, (ii) we keep track of
agents with monotonically increasing opinions as well as agents with
monotonically increasing opinions, and (iii) we rescale the opinion
$v\to ve^{1/4}$ once per unit time. The latter rescaling enables
direct measurement of $V=ve^{t/4}$, and additionally, it prevents
simulation of exponentially small opinions.

\section{Positivity}
\label{sec:positive}

\begin{figure}[t]
\begin{center}
\includegraphics[width=0.45\textwidth]{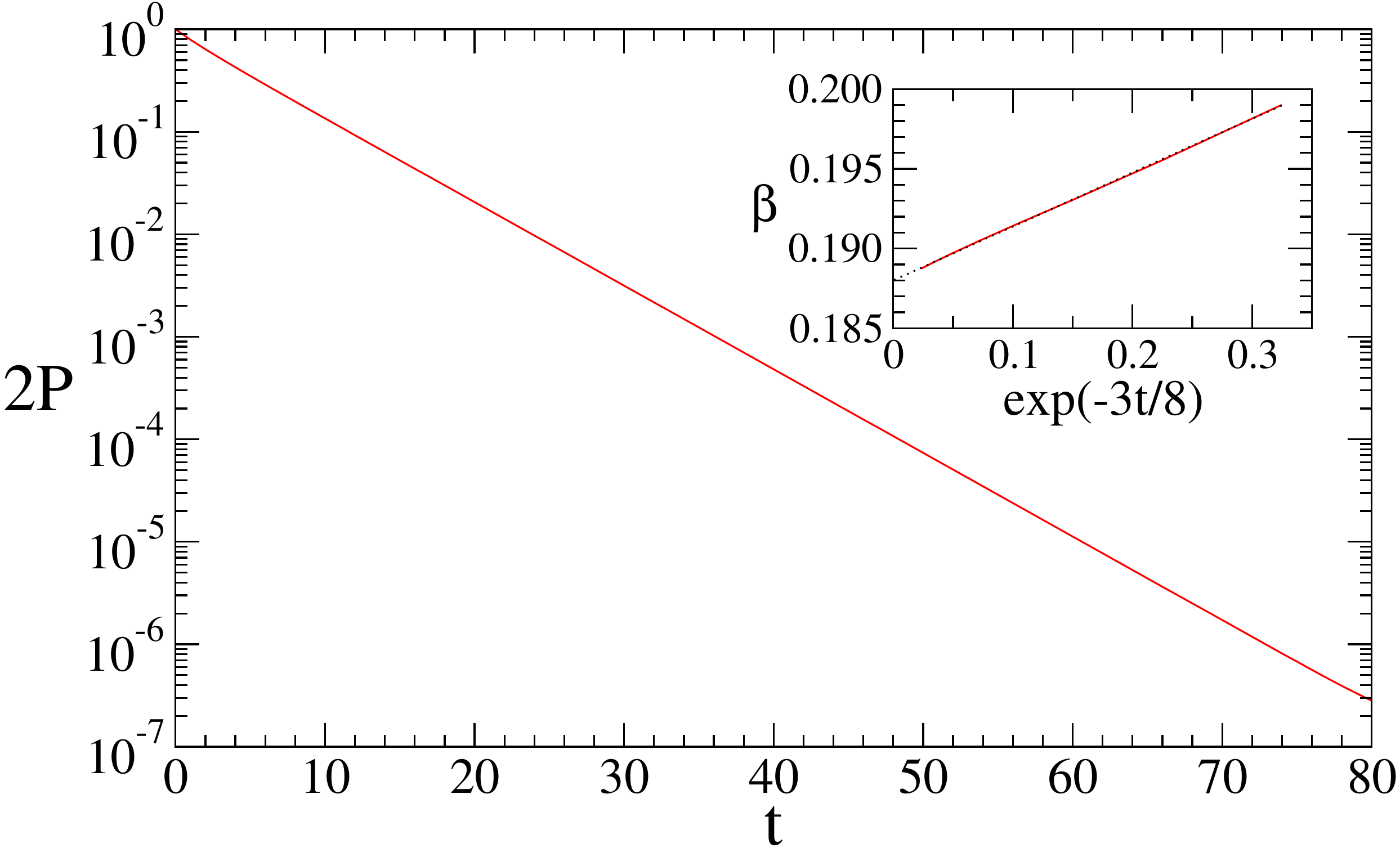}
\caption{The normalized fraction of positive agents $2P(t)$ versus
  time $t$ (since only one half of all agents start with a positive
  opinion, $2P(0)=1$). The inset shows $\beta(t) \equiv -d\ln P /\ln t$
  versus $\exp(-3t/8)$. A linear fit, shown as the dashed line,
  yields the estimate $\beta=0.188\pm 0.001$. }
\label{fig-Pt}
  \end{center}
\end{figure}

We also studied a related subset of agents with an opinion that
remains strictly above average. Since the average opinion
is zero, these are the agents with a strictly positive opinion.  Let
$P(t)$ be the fraction of such positive agents at time $t$.  Our
numerical simulations show that $P(t)$ decays exponentially with time
(Fig.~\ref{fig-Pt})
\begin{equation}
\label{Pt}
P(t)\simeq B\,e^{-\beta t}\quad {\rm with} \quad \beta=0.188\pm 0.01\,. 
\end{equation}

We analyze the density $P(v,t)$ of agents with opinion \hbox{$v>0$},
from which the overall density follows, \hbox{$P(t)=\int_0^\infty dv\,
P(v,t)$}.  Of course, there is a dual set of agents who maintain
a strictly negative opinion, with an  overall fraction $P(t)$ and
density $P(-v,t)$. In the context of opinion dynamics, positive and
negative agents can be viewed as agents that remain consistently on
the left side or the right side of the political spectrum.  The
density $P(v,t)$ evolves according to the {\em linear} equation
\begin{equation}
\label{Pvt-eq}
\frac{\partial P(v,t)}{\partial t} = 2\int_0^\infty du\,P(u,t)F(2v-u,t)-P(v,t)\,,
\end{equation}
for $v>0$. This equation reflects that the density $P(v,t)$ is coupled
to the overall density $F(v,t)$, and it differs from \eqref{Mvt-eq}
only in the lower limit of integration.  Using \eqref{Pvt-eq}, we
deduce the rate equation
\begin{equation}
\label{Pt-eq}
\frac{dP(t)}{dt} = - \int_0^\infty du\,P(u,t)\int_u^\infty dv F(v,t)
\end{equation}
for the fraction $P(t)$.  Again, this evolution equation differs from
\eqref{Mt-eq} only in the lower limit of integration.

\begin{figure}[t]
\begin{center}
\includegraphics[width=0.45\textwidth]{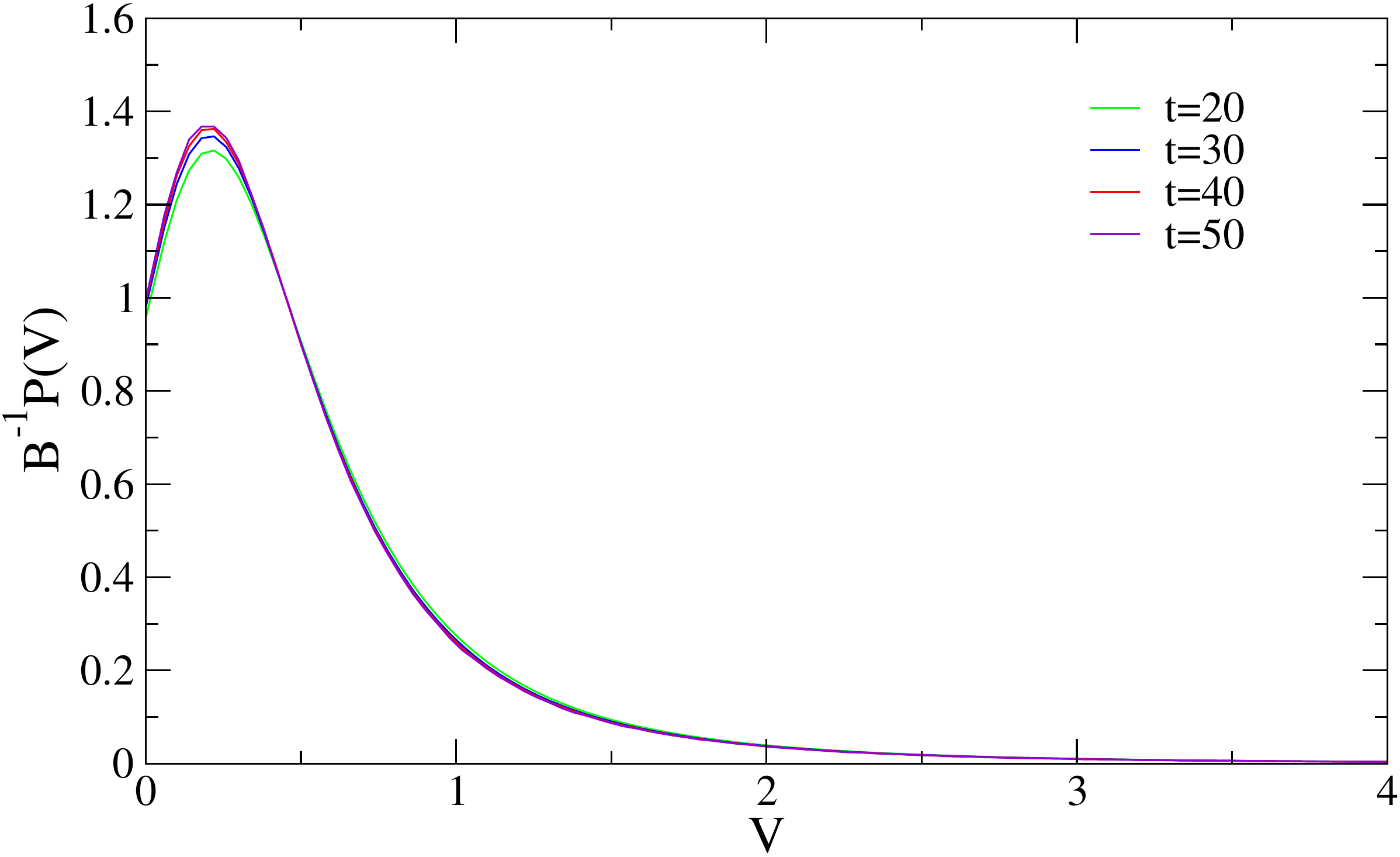}
\caption{The normalized scaling function $B^{-1}{\cal P}(V)$ versus
  the scaled opinion $V$.}
\label{fig-PV}
  \end{center}
\end{figure}

In the long-time limit, the distribution $P(v,t)$ acquires the scaling
form 
\begin{equation}
\label{Pvt-scaling}
P(v,t) \simeq e^{-\beta t}e^{t/4}\,\mathcal{P}(ve^{t/4})\,.
\end{equation}
This form is consistent with the exponential decay \eqref{Pt} with $B
= \int_0^\infty dV\,\mathcal{P}(V)$\,.  Our numerical simulations
confirm that this scaling behavior applies at sufficiently large
times (Fig.~\ref{fig-PV}). The scaling function $\mathcal{P}(V)$ is
qualitatively similar to $\mathcal{M}(V)$ (see Fig.~\ref{fig-MV}):
both functions are non-monotonic and are maximal at a nonzero value of
$V$. For positive agents, the ``depletion'' region near $V=0$ reflects
that positive agents with a sufficiently small opinion are less likely
to remain positive.

By substituting the scaling forms \eqref{Fvt-scaling} and
\eqref{Pvt-scaling} into the evolution equation \eqref{Pvt-eq}, we
find that the scaling function $\mathcal{P}(V)$ satisfies the linear
integro-differential equation
\begin{equation}
\label{PV-eq}
\frac{V}{4}\,\frac{d\mathcal{P}}{dV} + \left(\frac{5}{4}-\beta\right) \mathcal{P}
= 2\int_0^\infty dU\,\mathcal{P}(U)\,\mathcal{F}(2V-U).
\end{equation}
This equation poses an eigenvalue problem with the scaling function
$\mathcal{P}(V)$ being the eigenfunction, and the exponent $\beta$
being the eigenvalue. The eigenfunction is subject to the constraint
$\mathcal{P}(V)>0$ for all $V>0$.

\begin{figure}[t]
\begin{center}
\includegraphics[width=0.45\textwidth]{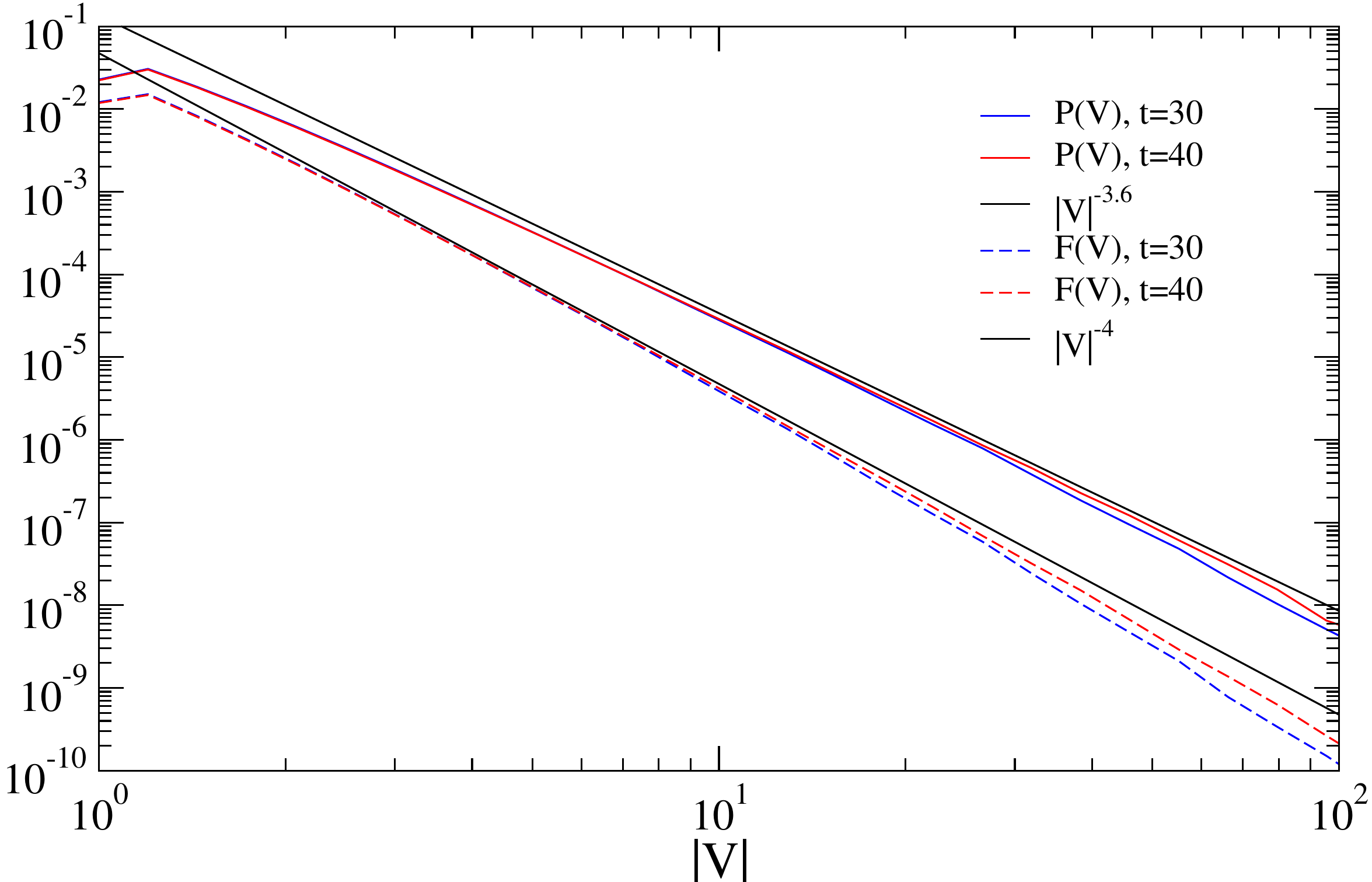}
\caption{The scaling function $\mathcal{F}(V)$ and $\mathcal{P}(V)$
  for $V\gg 1$. Also shown as a reference are the theoretical results
  \eqref{Fvt-scaling} and \eqref{beta-mu}.}
\label{fig-PV-tail}
  \end{center}
\end{figure}

To determine the asymptotic behavior of $\mathcal{P}(V)$ when $V\gg
1$, we repeat the approach used in Section \ref{sec:monotonic} and
arrive at an equation that is entirely analogous to \eqref{MV-eq},
\begin{equation}
\label{PV-asymp}
\frac{V}{4}\,\frac{d\mathcal{P}}{dV} + \left(\frac{5}{4}-\beta\right) \mathcal{P} = 
2\mathcal{P}(2V)\,.
\end{equation}
Hence, the tail of the scaling function is algebraic,
\hbox{$\mathcal{P}(V)\sim V^{-\mu}$}, and the dispersion relation reads
\begin{equation}
\label{beta-mu}
\beta=\frac{5-\mu}{4}-2^{1-\mu}\,.
\end{equation}
By substituting the Monte Carlo simulation result \eqref{Pt} into the
dispersion relation \eqref{beta-mu}, we expect \hbox{$\mu=3.58\pm
  0.01$}.  The numerical simulation results (Fig.~\ref{fig-PV-tail})
give \hbox{$\mu=3.6\pm 0.2$}, consistent with \eqref{beta-mu}.  The
eigenvalue $\beta$ does not correspond to an extremum in the
dispersion equation \eqref{beta-mu}, so it must be determined as the
eigenvalue of the full integro-differential equation \eqref{PV-eq}.
As was the case for monotonic agents, the large-$v$ tail of the
opinion density $P(v,t)$ is algebraic. Furthermore, the algebraic tail
of the scaling function $\mathcal{P}(V)$ is shallower than the tail of
the scaling function $\mathcal{F}(V)$ as $\mu<4$.

\section{Discussion}
\label{sec:summary}

In summary, we considered the averaging process and studied
sub-classes of agents with an opinion that maintains a certain
property throughout the evolution. In particular, we probed the
fraction of agents with a monotonically decreasing opinion and the
fraction of agents with a positive opinion. These fractions decrease
exponentially with time, and the exponents characterizing these decays
are eigenvalues of linear integro-differential equations.  In the case
of monotonic agents, we were able to find the eigenvalue analytically
using an extremum selection principle, analogous to velocity selection
in traveling waves
\cite{KPP,Fisher,Bramson,Saarloos,Saarloos-Rev,BD97,KM00,KM02,BK03,BD06}.
Both monotonicity and positivity can be viewed as types of
persistence, and as is typically the case, nontrivial persistence
exponents characterize the time evolution \cite{dhp,dbg,Bray13}.

For the averaging process, the opinion distribution becomes
self-similar, and in particular, the scaling form \eqref{Fvt-scaling}
is characterized by the second moment \eqref{M2}. However, moments of
the opinion distribution exhibit multiscaling and are not characterized
by the second moment. The moments decay exponentially with time
$\langle v^n\rangle\sim \exp(-\sigma_n t)$ with a nonlinear spectrum
of exponents, $\sigma_n/n\neq \sigma_2/2$ when $n>2$. We anticipate
that moments of the opinion distribution of monotonic agents also
exhibit multiscaling. Finding the corresponding
spectrum of exponents is challenging because the evolution equations
that govern the moments of $M(v,t)$ are not closed.

We also studied numerically the ultimate fraction $\rho$ of agents
with an opinion that obeys $|v(t)|\leq |v(0)|$ during the entire
evolution history, $0<t<\infty$. Interestingly, the fraction of such
agents is finite, and for a uniform distribution, the simulations
yield $\rho=0.6488\pm 0.0001$.  In contrast with the universal
exponents $\alpha$ and $\beta$, the fraction $\rho$ does depend on the
initial distribution.

A natural generalization of the averaging process is to
multi-component opinions.  When the opinion of each agent is a vector,
rather than a scalar, one may study monotonicity properties of the
magnitude of the opinion vector.  Additionally, one can investigate
agents with one component of the opinion vector being always larger
than all other components.

Monotonicity can also be studied in systems that reach a steady state,
and in particular, averaging processes that are forced into a steady
state \cite{bm}.  Monotonicity can be probed in an even broader class
of stochastic processes since the trajectory of any fluctuating
quantity may include segments where all changes in the value of the
fluctuating quantity occur in the same direction.

\acknowledgements

We dedicate this paper to Robert Ziff, whose singular style and
influential work continue to guide, inspire, and challenge an entire 
generation of statistical physicists.

\appendix

\section{Partial averaging}
\label{ap:partial}

In the partial averaging process, agents make a partial compromise by
moving part-way toward each other. The post interaction opinions
are linear combinations of the pre-interaction opinions
\begin{equation}
\label{partial}
(v_i,v_j) \to \left(pv_i+qv_j,qv_i+pv_j\right)\,,
\end{equation}
with $0\leq p \leq 1$ and $p+q=1$ so that the average opinion is
conserved.  Each interaction reduces the opinion difference by 
factor $|p-q|$, as in an inelastic collision \cite{BK00}.

Treatment of the partial averaging process is a straightforward
generalization of analysis above.  For the random process \eqref{partial}, the rate
equation governing the density $F(v,t)$ of agents with opinion $v$ at
time $t$ becomes
\begin{equation}
\label{Fvt-eq-p}
\frac{\partial F(v,t)}{\partial t} = 
\tfrac{1}{q}\int_{-\infty}^\infty\!\!\! du\,F(u,t)F\left(\tfrac{v-pu}{q},t\right)-F(v,t)\,.
\end{equation}
We can verify that the distribution remains normalized, $\int
dv\,F(v,t)=1$, and that the average opinion is conserved, $\int dv\,
v\, F(v,t)=0$. We restrict our attention to symmetric distributions,
and from \eqref{Fvt-eq-p}, it follows that the second moment decays
exponentially with time, $\langle v^2(t)\rangle = \langle
v^2(0)\rangle \, e^{-2pq t}$. In the long-time limit, the opinion
distribution follows the scaling form
\begin{equation}
\label{Fvt-scaling-p}
F(v,t) = e^{pq t}\mathcal{F}(V)
\end{equation}
with the scaling variable $V=ve^{pqt}$. Independent of $p$, the
scaling function $\mathcal{F}(V)$ is given by \eqref{Fvt-scaling}.

\begin{figure}[t]
\begin{center}
\includegraphics[width=0.45\textwidth]{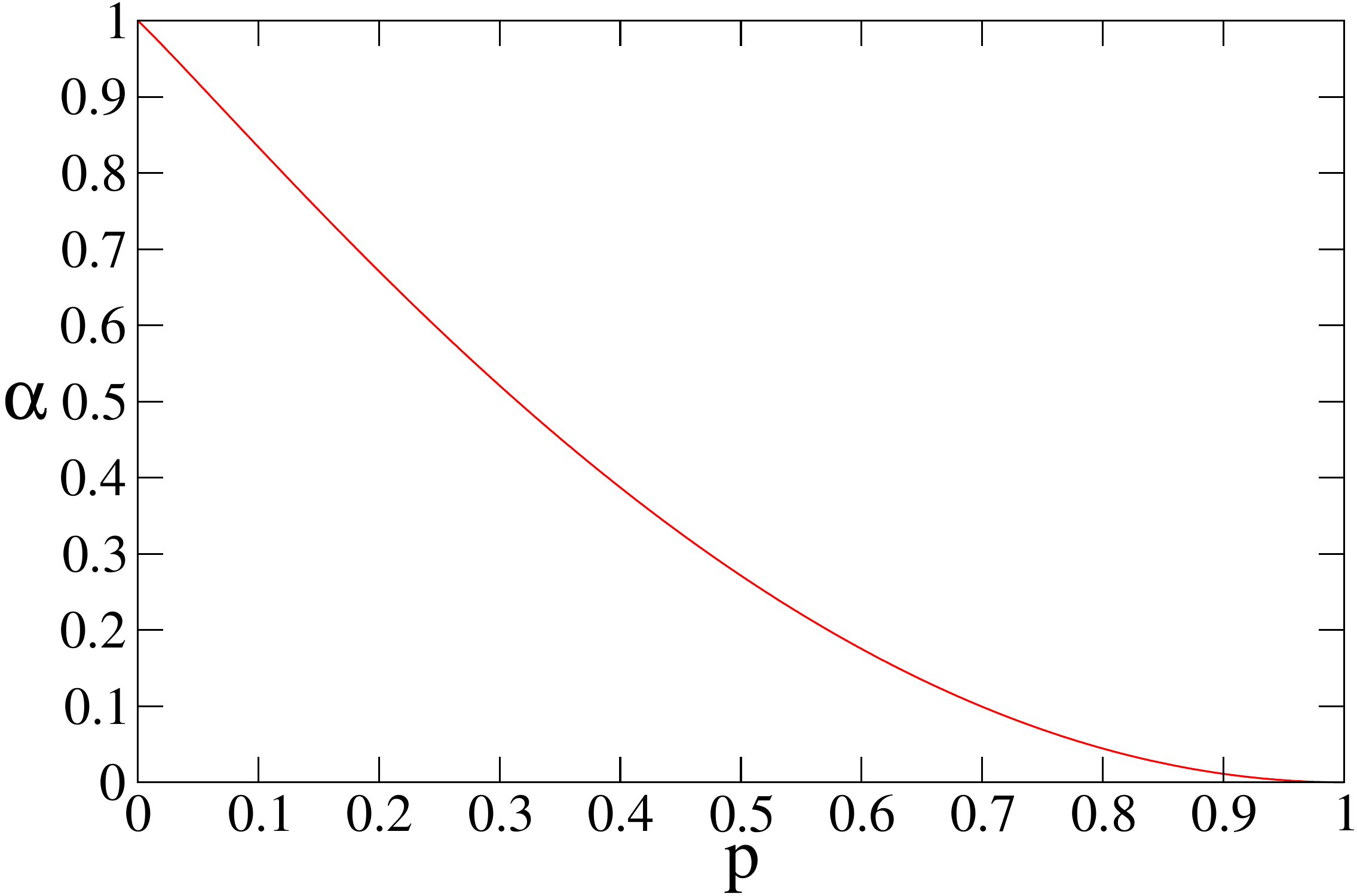}
\caption{The exponent $\alpha$, given by \eqref{alpha-max}, versus
  $p$. }
\label{fig:alpha-p}
\end{center}
\end{figure}

The density of monotonic agents $M(v,t)$ satisfies 
\begin{equation}
\label{Mvt-eq-p}
\frac{\partial M(v,t)}{\partial t}\!= \!\tfrac{1}{q}\int_v^\infty \!\!\!du\,M(u,t)F\left(\tfrac{v-pu}{q},t\right)-M(v,t)\,.
\end{equation}
In the long-time limit, the overall density of
monotonic agents decays as in \eqref{Mt}, and the density $M(v,t)$ of
monotonic agents approaches the scaling form \hbox{$M(v,t) \simeq
  e^{-\alpha t}e^{pq t}\,\mathcal{M}(V)$} with the scaling variable
\hbox{$V=ve^{pq t}$}.  The tail of the scaling function is algebraic,
$\mathcal{M}(V)\sim V^{-\nu}$ when \hbox{$V\gg 1$}, and equation
\eqref{alpha-nu} that relates the exponents $\alpha$ and $\mu$ becomes
\begin{equation}
\label{alpha-nu-p}
\alpha=1+pq(1-\nu)-p^{\nu-1}\,.
\end{equation}
The extremum  occurs at (figure
\ref{fig:alpha-p})
\begin{equation}
\label{alpha-max}
\alpha = 1 -pq \,\frac{1+\ln \left(\tfrac{1}{pq}\ln \tfrac{1}{p}\right)}{\ln \tfrac{1}{p}}\,.
\end{equation}
Our numerical simulations confirm that this extremum is indeed
selected by the dynamics.  The exponent \hbox{$0\leq \alpha \leq 1$}
decreases monotonically as $p$ increases. The maximal value $\alpha=1$
is achieved when the interaction is strongest ($p=0$), and the minimal
value $\alpha=0$ is achieved when the interaction is weakest
($p=1$). 

The exponent $\nu$ which characterizes the tail of the opinion distribution is given by 
\begin{equation}
\nu = 2 + \frac{\ln \left(\tfrac{1}{q}\ln \tfrac{1}{p}\right)}{\ln \tfrac{1}{p}}\,.
\end{equation}
The exponent $\nu$ increases monotonically with $p$. Since $\nu<4$,
the tail of the scaling distribution $\mathcal{M}(V)$ remains 
less steep than the tail of the scaling distribution $\mathcal{P}(V)$.

\section{Correction to the leading asymptotic behavior \eqref{Mt}}
\label{ap:corr}

In problems admitting traveling wave solutions, the speed $w$ and the
wavenumber $k$ of the traveling wave are related via the dispersion
relation $w=\Phi(k)$, and the
propagation velocity $w=w_*=\Phi(k_*)$ is selected at an extremum of
the dispersion curve.  This happens in a broad set of problems. 
Moreover, the asymptotic approach to the speed
$w_*$ is remarkably universal
\begin{equation}
\label{w-t}
w(t) = w_* + \frac{3}{2 k_*}\,t^{-1}+ B_2 t^{-3/2} + \cdots\,.
\end{equation}
The leading $t^{-1}$ correction was derived by Bramson \cite{Bramson}
in the context of the Fisher-Kolmogorov  equation
\cite{KPP,Fisher}, and then confirmed for many other deterministic
\cite{Saarloos,Saarloos-Rev} and stochastic
\cite{BD97,KM00,KM02,BK03,BD06} systems. The second 
$t^{-3/2}$ correction was established in Ref.~\cite{Saarloos}.

In our problem, the dispersion relation is given by
Eq.~\eqref{alpha-nu}. By assuming \eqref{w-t} is valid, the correction
to the leading asymptotic behavior \eqref{alpha} is given by
\begin{equation}
\label{alpha-t}
\alpha(t) = \alpha_* + \frac{3}{2 \nu_*}\,t^{-1}+ B_2 t^{-3/2}+\cdots \,.
\end{equation}
This form implies an algebraic correction to the leading asymptotic
behavior \eqref{Mt} as $M(t)\simeq A\,t^{-a}\exp(-\alpha t)$ with
$a=3/(2\nu_*)$.  The inset in Fig.~\ref{fig-Mt} shows that
\hbox{$\alpha(t) \equiv -d\ln M /\ln t$} varies linearly with the
inverse time $1/t$. A two-parameter linear fit, accounting only for
the leading correction to the asymptotic behavior, yields the value
$\alpha=0.272\pm 0.003$. However, equation \eqref{alpha-t} which
accounts for the two leading corrections also involves only two
parameters, $\alpha$ and $B_2$ because $\alpha$ and $\nu$ are related
by \eqref{alpha-nu}. By using \eqref{alpha-t} and \eqref{alpha-nu}, we
obtain the improved estimate $\alpha=0.2715\pm 0.0005$ from the very
same simulation results. Such an approach is in the spirit of Bob
Ziff's work on numerous problems including percolation
\cite{Ziff-perc}, planar dimer tilings \cite{Ziff-dimers}, and random
sequential adsorption \cite{Ziff-RSA}, where he perfected the art of
using finite-time and finite-size corrections for producing
high-precision measurements from Monte Carlo simulations.


\begin{thebibliography}{99}
\bibitem{finite} We assume that variance of the opinion distribution
  is finite initially.
\end{thebibliography}
\end{document}